\documentclass[a4paper,11pt]{article}
\pdfoutput=1
\usepackage{jheppub}

\usepackage{amssymb,amsmath,amsfonts}

\newcommand{\overbar}[1]{\mkern 1.5mu\overline{\mkern-2mu#1\mkern-1mu}\mkern 1.5mu}
\newcommand{\overbarl}[1]{\mkern 1.5mu\overline{\mkern-1.5mu#1\mkern+0.5mu}\mkern 1.5mu}
\newcommand{\overbarA}[1]{\mkern 4mu\overline{\mkern-4mu#1\mkern-1mu}\mkern 1.5mu}

\renewcommand{\O}[1]{\ensuremath{\mathcal{O}\left(#1\right)}}
\renewcommand{\L}{\ensuremath{\mathcal{L}}}
\renewcommand{\o}[1]{\ensuremath{o\left(#1\right)}}
\newcommand{\W}{\ensuremath{\mathcal{W}}}
\newcommand{\Lb}{\ensuremath{\overbar{\mathcal{L}}}}
\newcommand{\Wb}{\ensuremath{\overbar{\mathcal{W}}}}

\newcommand{\Q}{\ensuremath{\mathcal{Q}}}
\newcommand{\Qb}{\ensuremath{\overbar{\mathcal{Q}}}}
\newcommand{\Ab}{\ensuremath{\overbarA{A}}}
\newcommand{\ab}{\ensuremath{\bar{a}}}
\newcommand{\tr}{\ensuremath{\mathrm{tr}}}
\newcommand{\eb}{\ensuremath{\overbarl{\epsilon}}}
\newcommand{\veb}{\ensuremath{\overbarl{\varepsilon}}}

\newcommand{\gb}{\ensuremath{\overbar{g}}}
\newcommand{\eL}{\ensuremath{\epsilon_{L}}}
\newcommand{\eW}{\ensuremath{\epsilon_{W}}}
\newcommand{\cb}{\ensuremath{\overbarl{c}}}

\newcommand{\eLn}{\ensuremath{\epsilon_{Ln}}}
\newcommand{\eWn}{\ensuremath{\epsilon_{Wn}}}
\newcommand{\eLnn}[1]{\ensuremath{\epsilon_{L #1}}}
\newcommand{\eWnn}[1]{\ensuremath{\epsilon_{W #1}}}

\newcommand{\ebLn}{\ensuremath{\overbarl{\epsilon}_{Ln}}}
\newcommand{\ebWn}{\ensuremath{\overbarl{\epsilon}_{Wn}}}
\newcommand{\ebLnn}[1]{\ensuremath{\overbarl{\epsilon}_{L #1}}}
\newcommand{\ebWnn}[1]{\ensuremath{\overbarl{\epsilon}_{W #1}}}

\DeclareMathOperator{\extdm}{d}
\newcommand{\extd}{\extdm \!}
\newcommand{\Lt}{{\tt L}}
\newcommand{\Wt}{{\tt W}}
\newcommand{\R}{{\mathbb{R}}}

\title{Lifshitz Holography with Isotropic Scale Invariance}

\author[a]{Michael Gary}
\author[a]{Daniel Grumiller}
\author[a]{Stefan Prohazka}
\author[b]{Soo-Jong Rey}

\affiliation[a]{Institute for Theoretical Physics,
          Vienna University of Technology,\\
          Wiedner Hauptstrasse 8--10/136,
          A-1040 Vienna, \rm AUSTRIA}

\affiliation[b]{School of Physics \& Center for Theoretical Physics\\
          Seoul National University, Seoul 151-747, \rm KOREA}

\emailAdd{mgary@hep.itp.tuwien.ac.at}
\emailAdd{grumil@hep.itp.tuwien.ac.at}
\emailAdd{prohazka@hep.itp.tuwien.ac.at}
\emailAdd{sjrey@snu.ac.kr}

\abstract{
Is it possible for an anisotropic Lifshitz critical point to actually exhibit isotropic conformal invariance? We answer this question in the affirmative by constructing a concrete holographic realization. We study three-dimensional spin-3 higher-spin gauge theory with a $z=2$ Lifshitz ground state with non-trivial spin-3 background.
We provide consistent boundary conditions and determine the associated asymptotic symmetry algebra. Surprisingly, we find that the algebra consists of two copies of the $\W_3$ extended conformal algebra, which is the extended conformal algebra of an isotropic critical system. Moreover, the central charges are given by $3\ell/(2G)$. We consider the possible geometric interpretation of the theory in light of the higher spin gauge invariance and remark on the implications of the asymptotic symmetry analysis.
}

\keywords{Lifshitz critical system, higher spin gravity, AdS/CFT correspondence, gauge/gravity duality, Chern-Simons gravity, W-algebras, non-relativistic holography}

\preprint{TUW--14--01}

\begin{document}
\maketitle

\section{Introduction}\label{intro}

A variety of condensed matter systems exhibits anisotropic scaling near a renormalization group fixed point.  Classical Lifshitz fixed points, in which the system scales anisotropically in different spatial directions, are extensively explored. Quantum Lifshitz fixed points, in which time and space scale anisotropically,  with relative scaling ratio $z$, are particularly common in strongly correlated systems \cite{Grinstein:1981,Hornreich:1975zz,Rokhsar:1988zz,Henley:1997,Sachdev:1999,Ardonne:2004,Ghaemi:2005,Varma:1997,Si:2003,Sachdev:1996,Yang:2004}. When such systems possess a master field configuration with large central charge, holography allows for a description in terms of weakly coupled gravitational theories. Many-body field theories describing such anisotropic fixed points were proposed to be holographically dual to gravity in the background of Lifshitz geometries, where time and space scale asymptotically with the same ratio $z$ \cite{Kachru:2008yh}.

The spacetime scaling of anisotropic fixed points is generally very different from the spacetime scaling of isotropic fixed points.
Even if the UV theory at the energy scale of the band gap would have a plethora of anisotropic interactions and lattice field effects, they are typically irrelevant perturbations, and the theory flows in the IR to a theory with emergent space(time) isotropy. Lifshitz fixed points are exceptions to such a general rule of thumb. Here, we pose a somewhat different question: Can an anisotropic Lifshitz critical point actually exhibit isotropic conformal invariance? We believe the answer to this question bears interesting and important implications to the aforementioned many-body field theories and beyond. In this paper, we shall answer this question in the affirmative by constructing an explicit realization. We shall formulate our realization in terms of a holographic duality between a gravitational theory in $(2+1)$ dimensional Lifshitz space and a non-gravitational theory in $(1+1)$ dimensions at a quantum Lifshitz fixed point, which we will assume also possesses higher-spin conserved currents. In this case, the gravitational theory is a higher-spin extension of Lifshitz gravity in $(2+1)$ dimensions.

Higher spin gauge theories \cite{Fronsdal:1978rb,Fang:1978wz,Fradkin:1986qy,Fradkin:1986ka,Vasiliev:1986bq,Fradkin:1987ks,Fradkin:1987ah,Vasiliev:1990en,Vasiliev:1990vu,Vasiliev:1992av,Vasiliev:1995dn,Vasiliev:1999ba,Vasiliev:2003ev,Vasiliev:2004qz,Bekaert:2005vh}, long studied for their own unitary generalization of spin-2 gravity and for their intriguing relation to the high-energy and tensionless limits of string theory (for recent reviews, see \cite{Sagnotti:2011qp,Didenko:2014dwa} and the references therein), have more recently also gained much attention due to their special role in the holography of the Anti-de~Sitter/conformal field theory (AdS/CFT) correspondence: higher-spin gauge theories are the gravitational duals to conformal field theories at Gaussian or Wilson--Fisher-type fixed points \cite{Sezgin:2002rt,Klebanov:2002ja,Maldacena:2011jn,Maldacena:2012sf}. Higher spin gauge theories are particularly simple in $(2+1)$ dimensions and admit a Chern--Simons description  \cite{Blencowe:1988gj,Bergshoeff:1989ns}, along the same lines as the Chern--Simons description of $(2+1)$ dimensional Einstein gravity \cite{Achucarro:1987vz,Witten:1988hc}. There is a further, extra simplification in $(2+1)$ dimensions: higher spin theories admit a truncation to a finite tower of spins \cite{Aragone:1983sz}, which allows the use of methods very similar to those used in studying spin-2 Einstein gravity \cite{Henneaux:2010xg,Campoleoni:2010zq}.

As in their higher-dimensional counterparts, the most symmetric solution to $(2+1)$-dimensional higher-spin gauge theory is AdS space. The theory also admits non-AdS solutions \cite{Gary:2012ms}. In particular, these include solutions with asymptotic Lifshitz scaling, distinct from asymptotic AdS solutions with isotropic scaling. We shall adopt the procedure outlined in \cite{Afshar:2012nk} and study the higher-spin gauge theory that contains spin-2 and -3 fields only. We shall construct a Lifshitz spacetime with relative scaling ratio $z=2$. A novel feature of our construction is that, unlike other constructions known so far that involve $p$-form gauge fields, the Lifshitz spacetime is realized by turning on a non-trivial spin-3 gauge field configuration. To address the question we posed in the beginning, we study consistent boundary conditions that lead to finite, conserved, and integrable asymptotic charges. 
We analyze the asymptotic symmetry algebra (ASA) and 
identify by holography the current algebra of the dual $(1+1)$-dimensional quantum Lifshitz fixed point. Remarkably, we find that the ASA is $\W_3 \oplus \W_3$, the same ASA as for AdS. We take this as evidence that the current algebra of quantum Lifshitz fixed points is actually enhanced to that of isotropic fixed points (in so far as the fixed point admits a holographic description).

Recently, asymptotic Lifshitz spacetimes were also studied by Gutperle, Hijano and Samani in a spin-3 gravity context \cite{Gutperle:2013oxa}. Much of their work focused on other issues such as the presence of black hole states in higher-spin Lifshitz backgrounds, holonomy conditions for such black holes, and the realization of a Lifshitz sub-algebra within the ASA. Here, we study a different aspect of higher-spin realization of the Lifshitz background. We examine fully consistent boundary conditions, analyze the resulting ASA, and determine the enhanced current algebra of the Lifshitz fixed point. On the way, we pay special attention to the issue of gauge invariance and possible interpretation of higher spin excitations on a Lifshitz background as a geometric theory.

The outline of the paper is as follows. We begin by briefly reviewing asymptotic Lifshitz spacetimes in section \ref{Lifshitz}. Then, in section \ref{strict}, we propose a set of Lifshitz boundary conditions in $\mathfrak{sl}(3, \mathbb{R}) \oplus \mathfrak{sl}(3, \mathbb{R})$ spin-3 gravity. In section \ref{se:4} we prove the consistency of our boundary conditions and identify the resulting ASA as $\W_3\oplus\W_3$. In section \ref{se:5} we show that, despite the presence of AdS isometries, the ground state on the gravity side is the $z=2$ Lifshitz spacetime with non-trivial spin-3 field background. Finally, in section \ref{concl}, we conclude with some remarks on more general Lifshitz boundary conditions, aspects on Lifshitz CFT$_2$ duals, comparison with the approach of \cite{Gutperle:2013oxa}, and a discussion of the role of geometry and gauge invariance.

\section{Lifshitz Spacetime in $\boldsymbol{(2+1)}$ Dimensions}\label{Lifshitz}

The $(2+1)$-dimensional Lifshitz spacetime \cite{Kachru:2008yh} is described by the line element
\begin{equation}\label{lineElr}
\extd s_{\mathrm{Lif}_{z}}^2 = \ell^2 \big(-r^{2z}\,\extd t^2 + \frac{\extd r^2}{r^2} + r^2\,\extd x^2 \big)\, .
 \end{equation}
The Lifshitz spacetime (\ref{lineElr}) is invariant under the anisotropic scaling ($z\in\mathbb{R}$):
\begin{equation}
  \label{anisotrop}
  t \rightarrow \lambda^z t \qquad x \rightarrow \lambda x \qquad r\rightarrow \lambda^{-1} r \, .
\end{equation}
For $z=1$, the scaling is isotropic and the spacetime (\ref{lineElr}) reduces to Poincar\'e patch AdS$_3$.

It is often useful to consider a change of coordinates to the radial variable $\rho = \ln  r$.
The spacetime (\ref{lineElr}) now becomes
\begin{equation}\label{Lifshitz3}
\extd s_{\mathrm{Lif}_{z}}^2 = \ell^{2} \left(- e^{2z\rho}\extd t^2 + \extd\rho^2 + e^{2\rho}\extd x^2\right)\, .
\end{equation}
The asymptotic region is approached for $\rho \rightarrow \infty$.

The Lifshitz spacetime (\ref{Lifshitz3}) possesses spacetime isometries. These Lifshitz isometries are generated by the Killing vector fields
\begin{equation}
 \xi_{\mathbb{H}} = \partial_t \qquad
 \xi_{\mathbb{P}} = \partial_x \qquad
 \xi_{\mathbb{D}} = -zt\,\partial_t+\partial_\rho -x\,\partial_x
\label{eq:iso}
\end{equation}
whose isometry algebra is the Lifshitz algebra $\mathfrak{lif}(z,\mathbb{R})$
\begin{equation}
 [\xi_{\mathbb{H}},\,\xi_{\mathbb{P}}] = 0 \qquad [\xi_{\mathbb{D}},\,\xi_{\mathbb{H}}] = z\,\xi_{\mathbb{H}}\qquad [\xi_{\mathbb{D}},\,\xi_{\mathbb{P}}] = \xi_{\mathbb{P}}
\label{eq:Lalgebra}
\end{equation}
The Killing vector $\xi_{\mathbb{H}}$ ($\xi_{\mathbb{P}}$) [$\xi_{\mathbb{D}}$] generates time translations (spatial translations) [anisotropic dilatations]. The Lifshitz spacetime with $z=1$ corresponds to the Poincar\'e patch of the isotropic AdS$_3$ spacetime. With enhanced $(1+1)$-dimensional Lorentz (boost) invariance, the isometry algebra gets enlarged to $\mathfrak{sl}(2, \mathbb{R}) \oplus \mathfrak{sl}(2, \mathbb{R})$ associated with two copies of chiral and anti-chiral excitations. 
Conversely, the Lifshitz algebra $\mathfrak{lif}(1, \mathbb{R})$ is a subalgebra of the $\mathfrak{sl}(2, \mathbb{R}) \oplus \mathfrak{sl}(2, \mathbb{R})$ isometry algebra of the AdS$_3$ spacetime.

Since the Lifshitz spacetime does not fulfill the vacuum Einstein equations, matter contributions are necessary. Known realizations so far involve, e.g., $p$-form gauge fields \cite{Kachru:2008yh}. For example, AdS Einstein gravity coupled to two 1-form abelian gauge fields $F_2 = \extd A_1, \ G_2 = \extd C_1$,
\begin{equation}\label{quadratic}
I = \frac{1}{16\pi G_3} \int \extd^3 x \sqrt{-g} \left[ R(g) + \frac{2}{\ell^{2}} + \frac{1}{4} ||F_2||^2 + \frac{1}{4\alpha} ||G_{2}||^2 + \frac{1}{2} \ast( A_{1} \wedge G_{2}) \right]\,,
\end{equation}
admits the Lifshitz spacetime as a classical solution, where the scaling ratio $z$ is determined by
\begin{equation}
z = \alpha \pm \sqrt{\alpha^2-1} \qquad (\alpha \ge 1)\,.
\end{equation}
Some other constructions require either a massive gauge field \cite{Taylor:2008tg}, a massive graviton \cite{AyonBeato:2009nh,Lu:2012xu} or Ho\v{r}ava--Lifshitz gravity \cite{Griffin:2012qx}.

Here, we take a different route and realize the Lifshitz spacetime by coupling AdS$_3$ Einstein gravity to a spin-3 field with full higher-spin gauge symmetry. In the next section, we construct an explicit example of  $(2+1)$-dimensional $z=2$ Lifshitz spacetime \eqref{Lifshitz3} with non-trivial spin-3 background field. We shall then carefully examine boundary conditions for the gravitational and spin-3 excitations over this Lifshitz spacetime.

\section{Lifshitz Boundary Conditions in Higher-Spin Gravity}\label{strict}

In the classical regime we shall be working on $(2+1)$-dimensional spin-3 gravity, which is most conveniently described in the gauge theory formulation. The action is that of an $\mathfrak{sl}(3, \mathbb{R}) \oplus \mathfrak{sl}(3, \mathbb{R})$ Chern-Simons gauge theory on the spacetime ${\cal M}$,
\begin{equation}\label{action}
I_{\mathrm{bulk}} = \frac{k}{4\pi}\int_{\cal M} \tr\left[CS(A)-CS(\Ab)\right]\ ,
\end{equation}
where $A$ and $\Ab$ are $\mathfrak{sl}(3, \mathbb{R})$ connections,
\begin{equation}\label{CS}
CS(A) = A\wedge \extd A + \tfrac{2}{3} A\wedge A\wedge A
\end{equation}
is the Chern--Simons 3-form, and the Chern--Simons level $k$ is inversely related to the gravitational coupling $G_3$ according to the formula
\begin{equation}
  \label{ktoG}
  k=\frac{\ell}{8 G_{3} \tr(\Lt_{0} \Lt_{0})}\ .
\end{equation}
Hereafter, we consider the principal embedding $\mathfrak{sl}_2\hookrightarrow \mathfrak{sl}_3$ with spin-2 generators $\Lt_{-1}, \Lt_0, \Lt_{+1}$ and spin-3 generators $\Wt_{-2}, \Wt_{-1}, \Wt_0, \Wt_{+1}, \Wt_{+2}$. Our conventions for the generators and traces are summarized in appendix \ref{app:A}; in particular, $\tr(\Lt_{0}\Lt_{0})=2$.
The classical equations of motion derived from the action \eqref{action} imply gauge flatness of the connections:
\begin{equation}
\extd A + A\wedge A = 0 = \extd \Ab + \Ab \wedge \Ab
 \label{eq:eom}
\end{equation}

In order to find the Lifshitz spacetime, we decompose the connections as in \cite{Banados:1994tn,Campoleoni:2010zq,Afshar:2012nk},
\begin{equation}\label{decomposition}
A = b^{-1}\extd b + b^{-1}\left( \hat{a}^{(0)} + a^{(0)} + a^{(1)} \right)b\,,\quad
\Ab = b\extd b^{-1} + b\left( \hat{\ab}^{(0)} + \ab^{(0)} + \ab^{(1)} \right)b^{-1}
\end{equation}
with the group element $b=e^{\rho \Lt_0}$. The connection thus splits into a sum of terms containing $\hat{a}^{(0)}$ and $a^{(0)}$ of order $\O{1}$, and $a^{(1)}$ of order $\o{1}$ (and similarly for the barred sector)\footnote{%
Given a function $f(\rho)$ that depends on the radial coordinate $\rho$, the notation $f$ is of order ${\cal O}(1)$ means $\lim_{\rho\to\infty} |f| < \infty$, while $f$ is of order $o(1)$ means $\lim_{\rho\to\infty} f = 0$.}. The fixed background $\hat{a}^{(0)}$ and the state-dependent fluctuations $a^{(0)}$ are assumed to satisfy the equations of motion \eqref{eq:eom}.
To fix a variational principle, we take $\delta A_t=0=\delta\Ab_{t}$ at asymptotic infinity $\rho \rightarrow \infty$, where $t$ and $x$ are our boundary coordinates. With the boundary term
\begin{equation}
\frac{k}{4\pi}\int_{\partial\mathcal{M}}\!\!\!\tr\left(  A_{t} A_{x} - \Ab_{t} \Ab_{x} \right) \extd t  \extd x
\label{eq:bt}
\end{equation}
added to the bulk action (\ref{action}), such a variational principle is well-posed \cite{Gary:2012ms}.
The boundary term (\ref{eq:bt}) retains the time-reversal ${\cal T}$ of the action (\ref{action}), under which the temporal orientation of ${\cal M}$ is changed, $t\to-t$, together with $A\to-\bar A$, $\bar A\to -A$, $\Lt_n\to \Lt_{-n}$ and $\Wt_n\to \Wt_{-n}$.

We take as a background that leads to the Lifshitz spacetime the connections
\begin{subequations}
\label{background}
  \begin{align}
    \hat{a}^{(0)} &=\tfrac{4}{9}\Wt_{+2}\extd t +  \Lt_{+1}\extd x \\
    \hat{\ab}^{(0)} &= \Wt_{-2}\extd t + \Lt_{-1}\extd x \, .
  \end{align}
  \end{subequations}
The specific numerical coefficients are chosen to cancel factors arising from traces. Note that the background breaks the time-reversal ${\cal T}$. 

Using the standard definition of the metric in terms of the zuvielbein,
\begin{align}\label{metricDef}
g_{\mu\nu} &= \frac{1}{2}\tr\left(e_\mu e_\nu\right) \quad \mbox{where} \quad    e_\mu = \frac{\ell}{2}\left(A_\mu - \Ab_\mu\right),
\end{align}
leads to the geometry
\begin{equation}\label{backgroundLineEl}
\extd s_{\mathrm{Lif_{2}}}^2 =\ell^2 \left(- e^{4\rho}\extd t^2+\extd\rho^2  + e^{2\rho}\extd x^2 \right)\, .
\end{equation}
We thus obtain as a classical configuration the $(2+1)$-dimensional Lifshitz spacetime \eqref{Lifshitz3} with $z=2$. The classical solution also involves the totally symmetric spin-3 gauge field:
\begin{equation}
  \label{spin3field}
\phi_{\lambda \mu \nu}=\frac{1}{3 !} \tr(e_{(\lambda}e_{\mu}e_{\nu)})\ ,
\end{equation}
where the parentheses denote symmetrization without further normalization. For our classical configuration, we find that the Lifshitz spacetime is supported by a nontrivial  spin-3 background gauge field
\begin{equation}\label{backgroundSpin3}
\phi_{\mu\nu\lambda}\,\extd x^\mu\extd x^\nu\extd x^\lambda = -\frac{5 \ell^{3}}{4}\, e^{4\rho}\,\extd t\,(\extd x)^2 \, .
\end{equation}
From now on we set $\ell=1$ to reduce clutter. The spin-3 gauge field is invariant under the transformations generated by the Killing vector fields (\ref{eq:iso}). We conclude that the classical configuration (\ref{backgroundLineEl}),  (\ref{backgroundSpin3}) respects the Lifshitz algebra $\mathfrak{lif}(2, \mathbb{R})$.

The above construction of the Lifshitz spacetime is quite elementary and simple. A potential advantage of the higher-spin realization of the Lifshitz spacetime is better control of stability due to enlarged gauge symmetry. It is believed (though precise details are yet to be understood better) that higher-spin gauge theory is a consistent unitary truncation when arising from compactification of a higher-dimensional UV completion such as string theory. If this were true, the higher-spin gauge symmetry severely constrains nonlinear interactions. For example, in the $p$-form construction, the full-fledged dynamics would include nonlinear interactions beyond the quadratic terms \eqref{quadratic} and other spectator fields must also have quadratic or higher-order interactions with the $p$-form fields. This is in general not automatic and neutral scalar fields, if present, are the most delicate ones to control. In contrast, the higher-spin theory does not face such issues since scalars that would arise from compactification are necessarily all charged under the higher-spin gauge symmetry and therefore severely constrained.


Let us next examine the algebra of the symmetry currents for the Lifshitz system we have constructed. To this end, we first need to impose boundary conditions consistent with the background Lifshitz spacetime geometry. Note that we take the ansatz used in \cite{Afshar:2012nk,Prohazka:2013}, which differs from the asymptotic behavior $A-\hat{A}=\O{1}$ used in \cite{Campoleoni:2010zq,Gutperle:2013oxa}, where $\hat{A}$ was a fixed background connection. The fluctuations, which are already on-shell, turn out to take the following form
\begin{subequations}
\label{strictBC}
  \begin{align}
    a^{(0)} &= \big(\tfrac{8\pi}{9k}t\W(x)\Lt_0 - \tfrac{\pi}{2k}\L(x)\Lt_{-1}\big)\,\extd x\nonumber\\
    &\quad + \big(-\tfrac{32\pi}{81k}t^2\W(x)\Wt_{+2} + \tfrac{8\pi}{9k}t\L(x)\Wt_{+1} + \tfrac{2\pi}{9k}\W(x)\Wt_{-2}\big)\,\extd x\label{strictBC1}\\
    \ab^{(0)} &= \big(- \tfrac{2\pi}{k}t\Wb(x)\Lt_0-\tfrac{\pi}{2k}\Lb(x)\Lt_{+1}\big)\,\extd x\nonumber\\
    &\quad + \big(- \tfrac{2\pi}{k}t^2\Wb(x)\Wt_{-2} - \tfrac{2\pi}{k}t\Lb(x)\Wt_{-1} +\tfrac{2\pi}{9k}\Wb(x)\Wt_{+2}\big)\,\extd x\label{strictBC3}\\
    a^{(1)} &= \o{1} = \ab^{(1)}\, .\label{strictBC4}
  \end{align}
\end{subequations}
The set of all boundary functions $\L$, $\Lb$, $\W$ and $\Wb$ specify the set of all admissible fluctuations about the Lifshitz background.
 
A remarkable feature of these boundary conditions is the polynomial time dependence. In general, time-dependent boundary conditions lead to non-conservation of canonical charges. However, in the next section, we will demonstrate that all $t$-dependence is canceled in the boundary charge density and hence the canonical charges are conserved.

Below, we address some immediate consequences of the above boundary conditions (\ref{strictBC}), which all point to the fact that consistency of the boundary conditions is a highly non-trivial result.

Using \eqref{metricDef} and \eqref{spin3field}, we also extract fluctuations of spin-2 and spin-3 fields. Up to the sub-leading terms $a^{(1)}$ and $\ab^{(1)}$, fluctuations of the spin-2 field take the form (for notational simplification, we suppress the $x$-dependence of all component functions hereafter)
\begin{subequations}
    \label{metricfluc}
  \begin{align}
    g_{t t}&= \boldsymbol{-e^{4 \rho }} \\
    g_{t \rho}&=0 \\
    g_{t x}&=t^2 e^{4 \rho } \big(\pi \Wb+\tfrac{4\pi}{9} \W\big)+\tfrac{\pi}{4}  \Wb+\tfrac{\pi}{9} \W \\
    g_{\rho \rho}&=\boldsymbol{1} \\
    g_{\rho x}&=t\,\big(\tfrac{\pi}{2} \Wb+\tfrac{2\pi}{9} \W\big) \\
    g_{x x}&= \boldsymbol{e^{2 \rho }}- t^4e^{4 \rho }\tfrac{16\pi^2}{81} \Wb \W - t^2e^{2 \rho }\tfrac{\pi^2}{9} \Lb \L \nonumber\\
    &\quad+\tfrac{\pi}{6} \Lb+\tfrac{\pi}{6} \L+t^2\tfrac{8\pi^2}{81}
     \Wb \W+\tfrac{\pi^2}{36} e^{-2 \rho } \Lb \L-\tfrac{\pi^2}{81}
     e^{-4 \rho } \Wb \W \ ,
  \end{align}
\end{subequations}
while fluctuations of the spin-3 field take the form
\begin{subequations}
    \label{spin3fluc}
  \begin{align}
    \phi_{txx}&=\boldsymbol{-\tfrac{5}{12}e^{4 \rho }}+ t^2e^{4 \rho }\, \big(\tfrac{\pi ^2}{3 k^2}\L^2-\tfrac{3 \pi ^2}{4 k^2}\Lb^2\big)+e^{2 \rho}\, \big(\tfrac{\pi}{3 k}\L-\tfrac{3 \pi}{4 k}\Lb\big)+\tfrac{\pi ^2}{12 k^2}\L^2-\tfrac{3 \pi ^2}{16 k^2}\Lb^2 \\
    \phi_{\rho xx}&=te^{2 \rho }\, \big(\tfrac{2 \pi}{3 k}\L-\tfrac{3 \pi}{2 k}\Lb\big) + t\,\big(\tfrac{\pi ^2}{3 k^2}\L^2 - \tfrac{3 \pi ^2}{4 k^2}\Lb^2\big) \\
\nonumber    \\
    \phi_{xxx}&=t^4 e^{4 \rho }\, \big(\tfrac{2 \pi ^3}{k^3}\Lb^2 \W-\tfrac{2 \pi ^3}{k^3}\L^2 \Wb\big)+t^2 e^{4 \rho }\, \big(\tfrac{9 \pi}{2 k}\Wb-\tfrac{8 \pi }{9 k}\W\big)\nonumber\\
    &\quad+t^2 e^{2 \rho }\, \big(\tfrac{2 \pi ^2}{k^2} \L \Wb-\tfrac{2 \pi ^2 }{k^2}\Lb \W\big)+t^2\, \big(\tfrac{\pi^3}{k^3}\L^2 \Wb-\tfrac{\pi ^3  }{k^3}\Lb^2 \W\big)-\tfrac{\pi}{2 k}\Wb+\tfrac{\pi}{2 k} \W \nonumber\\
    &\quad+e^{-2 \rho }\, \big(\tfrac{\pi^2}{2 k^2}\Lb \W-\tfrac{\pi^2}{2 k^2}\L \Wb\big)+e^{-4 \rho }\, \big(\tfrac{\pi^3}{8 k^3}\Lb^2\W-\tfrac{\pi^3}{8 k^3}\L^2 \Wb\big) \\
    \phi_{\mu\nu\lambda}&= 0 \quad \textrm{otherwise}\, .
  \end{align}
\end{subequations}
The boldfaced terms denote background geometry, while the remaining terms correspond to state-dependent contributions to the spin-2 and spin-3 fields.

Note that the spin-3 field explicitly breaks the time-reversal symmetry ${\cal T}$. In particular, there are no states in the theory which are time-reversal invariant. This should be contrasted with AdS$_3$, which admits an infinite family of static solutions, all of which are time-reversal invariant. 

It is also interesting to observe that, although the background geometry is Lifshitz, the boundary conditions also admit spin-2 field configurations that have asymptotically stronger divergent contributions in $\rho$ than the background geometry. For example, it is possible to have configurations whose $g_{tt}$ and $g_{xx}$ have the same asymptotic growth, $\sim e^{4\rho}$. Nevertheless, as we are going to show below, all the configurations allowed by our boundary conditions correspond to finite energy excitations, in the sense that all the canonical charges associated with these configurations are finite (as well as integrable and conserved).  It should be stressed that this feature crucially relies on higher-spin gauge symmetry that acts nontrivially on the spin-2 metric field: the would-be infinite energy density in Einstein-gravity for configurations of $\sim e^{4 \rho}$ asymptotic growth is canceled off by the spin-3 gauge transformations in higher-spin gravity.

\section{Asymptotic Symmetry Algebra and Canonical Charges}\label{se:4}

We now examine the ASA of the higher-spin theory and its canonical charges. The bulk action \eqref{action} has gauge invariance of the form
\begin{align}
\delta_\epsilon A &= \extd\epsilon + \left[A,\epsilon\right] & \delta_{\eb}\Ab &= \extd\eb + \left[\Ab,\eb\right]\ .
\end{align}
Taking into account the decomposition \eqref{decomposition} and the boundary conditions \eqref{strictBC}, the gauge transformations that retain the boundary conditions are of the form
\begin{subequations}
\label{eq:lalapetz}
  \begin{align}
    \epsilon &= b^{-1}\left[\varepsilon_L + \varepsilon_W + \o{1}\right] b\\
    \eb &= b 
\left[\veb_L + \veb_W + \o{1}\right]b^{-1} \ .
  \end{align}
\end{subequations}
The rotated gauge functions $\varepsilon_L$, $\veb_L$, $\varepsilon_W$ and $\veb_W$ are further decomposable into fluctuation-independent parts labeled by 
$\epsilon_L$, $\eb_L$, $\epsilon_W$, $\eb_W$, and fluctuation-dependent parts proportional to ${\cal L}$, $\overline{\cal L}$, ${\cal W}$, $\overline{\cal W}$ 
 (for notational simplicity, we suppress the $x$-dependence of all the functions involved):
\begin{subequations}
\label{epsilonForm}
  \begin{align}
    \varepsilon_L &= \epsilon_L \Lt_{+1} + \big(\tfrac{8\pi}{9k}t\W\epsilon_L-\epsilon_L'\big)\, \Lt_0 - \big(\tfrac{\pi}{2k}\L\epsilon_L - \tfrac{1}{2}\epsilon_L''\big)\, \Lt_{-1}\nonumber\\
    &\quad - \big(\tfrac{32\pi}{81k}t^2\W\epsilon_L - \tfrac{8}{9}t\epsilon_L'\big)\, \Wt_{+2} + \big(\tfrac{8\pi}{9k}t\L\epsilon_L - \tfrac{8}{9}t\epsilon_L''\big)\, \Wt_{+1} + \tfrac{2\pi}{9k}\W\epsilon_L \Wt_{-2}\displaybreak[0]\\
\nonumber    \\
    \varepsilon_W &= \big(\tfrac{\pi}{3k}t\L'\epsilon_W + \tfrac{5\pi}{6k}t\L\epsilon_W' - \tfrac{1}{6}t\epsilon_W^{(3)}\big)\, \Lt_{+1}\nonumber\\
    &\quad + \big(\tfrac{\pi^2}{k^2}t\L^2\epsilon_W - \tfrac{\pi}{3k}t\L''\epsilon_W -\tfrac{7\pi}{6k}t\L'\epsilon_W' - \tfrac{4\pi}{3k}t\L\epsilon_W'' + \tfrac{1}{6}t\epsilon_W^{(4)}\big)\, \Lt_0- \tfrac{\pi}{k}\W\epsilon_W \Lt_{-1} \nonumber\\
    &\quad + \big(\epsilon_W -\tfrac{4\pi^2}{9k^2}t^2\L^2\epsilon_W + \tfrac{4\pi}{27k}t^2\L''\epsilon_W + \tfrac{14\pi}{27k}t^2\L'\epsilon_W' +\tfrac{16\pi}{27k}t^2\L\epsilon_W'' - \tfrac{2}{27}t^2\epsilon_W^{(4)}  \big)\,
    \Wt_{+2}\nonumber\\
    &\quad  + \big(\tfrac{16\pi}{9k}t\W\epsilon_W - \epsilon_W' \big)\, \Wt_{+1} - \big(\tfrac{\pi}{k}\L\epsilon_W - \tfrac{1}{2}\epsilon_W''\big)\,\Wt_0
\nonumber\\ &\quad
    + \big(\tfrac{\pi}{3k}\L'\epsilon_W + \tfrac{5\pi}{6k}\L\epsilon_W' - \tfrac{1}{6}\epsilon_W^{(3)}\big)\,\Wt_{-1}\nonumber\\
    &\quad + \big(
      \tfrac{\pi^2}{4k^2}\L^2\epsilon_W - \tfrac{\pi}{12k}\L''\epsilon_W - \tfrac{7\pi}{24k}\L'\epsilon_W'
      - \tfrac{\pi}{3k}\L\epsilon_W'' +
      \tfrac{1}{24}\epsilon_W^{(4)}\big)\,\Wt_{-2}
  \end{align}
and
  \begin{align}
    \veb_L &= \eb_L \Lt_{-1} - \big(\tfrac{2\pi}{k}t\Wb\eb_L - \eb_L'\big)\, \Lt_0 - \big(\tfrac{\pi}{2k}\Lb\eb_L - \tfrac{1}{2}\eb_L''\big)\, \Lt_{+1} \nonumber\\
    &\quad - \big(\tfrac{2\pi}{k}t^2\Wb\eb_L - 2t\eb_L'\big)\,\Wt_{-2} - \big(\tfrac{2\pi}{k}t\Lb\eb_L - 2t\eb_{L}''\big)\,\Wt_{-1} + \tfrac{2\pi}{9k}\Wb\eb_L \Wt_{+2}\displaybreak[0]\\
\nonumber    \\
    \veb_W &= \big(\tfrac{3\pi}{4k}t\Lb'\eb_W + \tfrac{15\pi}{8k}t\Lb\eb_W' - \tfrac{3}{8}t\eb_W^{(3)}\big)\, \Lt_{-1} \nonumber\\
    &\quad + \big(- \tfrac{9\pi^2}{4k^2}t\Lb^2\eb_W + \tfrac{3\pi}{4k}t\Lb''\eb_W +\tfrac{21\pi}{8k}t\Lb'\eb_W' + \tfrac{3\pi}{k}t\Lb\eb_W'' - \tfrac{3}{8}t\eb_W^{(4)}\big)\,\Lt_0 -\tfrac{\pi}{k}\Wb\eb_W \Lt_{+1} \nonumber\\
    &\quad + \big(\eb_W -\tfrac{9\pi^2}{4k^2}t^2\Lb^2\eb_W + \tfrac{3\pi}{4k}t^2\Lb''\eb_W + \tfrac{21\pi}{8k}t^2\Lb'\eb_W' + \tfrac{3\pi}{k}t^2\Lb\eb_W'' -
      \tfrac{3}{8}t^2\eb_W^{(4)} \big)\,\Wt_{-2} \nonumber\\
    &\quad - \big(\tfrac{4\pi}{k}t\Wb\eb_W - \eb_W'\big)\,\Wt_{-1}
    - \big(\tfrac{\pi}{k}\Lb\eb_W - \tfrac{1}{2}\eb_W''\big)\,\Wt_0  \nonumber\\
&\quad - \big(\tfrac{\pi}{3k}\Lb'\eb_W + \tfrac{5\pi}{6k}\Lb\eb_W' - \tfrac{1}{6}\eb_W^{(3)}\big)\,\Wt_{+1}\nonumber\\
    &\quad +\big(\tfrac{\pi^2}{4k^2}\Lb^2\eb_W - \tfrac{\pi}{12k}\Lb''\eb_W - \tfrac{7\pi}{24k}\Lb'\eb_W' - \tfrac{\pi}{3k}\Lb\eb_W'' + \tfrac{1}{24}\eb_W^{(4)}\big)\,\Wt_{+2} \, .\label{ebForm}
  \end{align}
\end{subequations}

The equations of motion assert that the canonical charges are given by the magnetic field flux. Inserting the results above into the variations of the canonical charges (see e.g.~\cite{Blagojevic}),
\begin{equation}
\delta \Q = \frac{k}{2\pi}\, \int \extd x \,\tr\big(\epsilon\, \delta A_x \big)\qquad \delta \Qb = \frac{k}{2\pi}\, \int \extd x \,\tr\big(\eb\, \delta \Ab_x \big)
 \label{eq:charges}
\end{equation}
one finds two non-trivial, state-dependent, finite, conserved, integrable canonical asymptotic charges in each of the unbarred and barred sectors:
\begin{subequations}\label{charges}
  \begin{align}
    \Q &= \int \extd x\left(\L(x)\epsilon_L(x) + \W(x)\epsilon_W(x)\right)\, ,\\
    \Qb &= \int \extd x\left(\Lb(x)\eb_L(x) + \Wb(x)\eb_W(x)\right)\, .
  \end{align}
\end{subequations}
Note that the charge densities depend linearly on the fluctuation-independent part and the fluctuation-dependent part, respectively. Note also that the charge densities are time-independent despite the situation that the boundary conditions \eqref{strictBC} are explicitly time-dependent. Typically, the Hamiltonian is part of the symmetry algebra and transforms covariantly under the algebra. This renders the canonical charges generically time-dependent. The Noether charges are a refinement of these canonical charges in which all time-dependences are removed by linear transformation at each step of time evolution. In this regard, our situation is exceptional since, for time-dependent boundary conditions, both the 
canonical charges and the Noether charges are time-independent. 
 
For completeness, one also needs to specify the global structure of $x$-space. We come back to this issue in section \ref{se:5}.
The canonical charges \eqref{charges} also prove that our boundary conditions \eqref{background}, \eqref{strictBC} are indeed consistent.

Using the shortcut (see e.g.~\cite{Henneaux:2010xg,Campoleoni:2010zq}),
\begin{equation}\label{shortcut}
\left\{\Q\left[\epsilon\right],\, \bullet\right\} := -\delta_\epsilon(\bullet)\ ,
\end{equation}
we can determine the ASA by evaluating the variations of the charges. For these variations, we find the result for the first, unbarred $\mathfrak{sl}(3, \mathbb{R})$ gauge connection
\begin{subequations}
\label{eq:angelinajolie}
  \begin{flalign}
    \delta_{\epsilon_L}\L &= \L'\epsilon_L + 2\L\epsilon_L' - \tfrac{k}{\pi}\epsilon_L^{(3)}&\\
    \delta_{\epsilon_L}\W &= \W'\epsilon_L + 3\W\epsilon_L'&\\
    \delta_{\epsilon_W}\L &= 2\W'\epsilon_W + 3\W\epsilon_W'&\\
    \delta_{\epsilon_W}\W &= \big(\tfrac{3\pi}{k}\L\L' - \tfrac{3}{8}\L^{(3)}\big)\epsilon_W + \big(\tfrac{3\pi}{k}\L^2 - \tfrac{27}{16}\L''\big)\epsilon_W'
- \tfrac{45}{16}\L'\epsilon_W'' -
    \tfrac{15}{8}\L\epsilon_W^{(3)} + \tfrac{3k}{16\pi}\epsilon_W^{(5)}.&
  \end{flalign}
\end{subequations}
Here, the superscript $(n)$ refers to $n$-th derivative with respect to the argument of the functions. 

Interestingly, we also find identical expressions for the second, barred $\mathfrak{sl}(3, \mathbb{R})$ gauge connection with the replacement $\epsilon \to \eb$, $\mathcal{L} \to \Lb$ and $\mathcal{W} \to \Wb$. This means that, although the background, boundary conditions and the admissible gauge functions are asymmetric between the unbarred and the barred sectors, the resulting ASA, and hence the canonical charges, exhibit exchange symmetry under the time-reversal ${\cal T}$. 
We recognize the ASA, \eqref{eq:angelinajolie} plus its barred sector, as the classical $\W_3\oplus\W_3$ extended conformal algebra (see e.g.~\cite{Blumenhagen} and appendix \ref{app:B}). It is remarkable that we have two identical copies of the $\W_3$ extended conformal algebra despite the fact that the Lifshitz spacetime lacks  $(1+1)$-dimensional Lorentz or boost invariance as well as the time-reversal symmetry. Recalling the relation between level $k$, Newton constant $G_3$, and curvature radius $\ell$, \eqref{ktoG}, as well as the conventions for the generator $\Lt_0$, the central charges of the extended conformal algebra reads
\begin{equation}\label{strictCentralCharges}
c = \cb = 12k \,\tr\left(\Lt_0\Lt_0\right) = 24k = \frac{3\ell}{2G_3}\, .
\end{equation}
The same value of the central charge follows for the  AdS$_3$ Einstein gravity \cite{Brown:1986nw} as well as AdS$_3$ higher-spin gravity~\cite{Henneaux:2010xg,Campoleoni:2010zq}, presumably reflecting the fact that all these theories do not have propagating bulk degrees of freedom.

\section{AdS Isometries of the Lifshitz Background}\label{se:5}

As stressed, the same ASA, $\W_3\oplus\W_3$ extended conformal algebra, with precisely the central charges \eqref{strictCentralCharges}, arose in spin-3 gravity with the isotropic, AdS$_3$ boundary conditions \cite{Henneaux:2010xg, Campoleoni:2010zq}. Thus, it is imperative to verify that our boundary conditions are not merely a complicated alternative route for imposing the same asymptotic AdS$_3$ boundary conditions. To check this, we now examine whether we have correctly identified the ground state of the theory.

In particular, if we have identified the holographic dictionary between higher-spin gravity and CFT correctly, the symmetries of individual states should agree on both sides of the duality. On the CFT side, Ward identities relate symmetries of a state to generators which annihilate the state. On the bulk side, such symmetries are given by gauge transformations which leave the state invariant and, adapting the language of gravity, are called isometries. It is important not to confuse isometries, which are symmetries of an individual state in a gravitational theory, with the symmetries of the theory as a whole, which often relate distinct states. In particular, the number of isometries of any particular state is always finite, while the symmetries of a theory can be much bigger and could (and in our case do) enhance \'a la Brown--Henneaux to infinite dimensional algebras.

The ground state of the dual CFT$_2$ ought to be invariant under the wedge algebra, $\mathfrak{sl}(3,\R)\oplus\mathfrak{sl}(3,\R)$, and thus annihilated by the generators $\L_{0},\L_{\pm1},\W_0,\W_{\pm1},\W_{\pm2}$ and their barred counterparts. For consistency, the Lifshitz ground state \eqref{backgroundLineEl}, \eqref{backgroundSpin3} on the higher-spin gravity side must also be $\mathfrak{sl}(3,\R)\oplus\mathfrak{sl}(3,\R)$ invariant.

The gauge variations that leave \eqref{background} and \eqref{strictBC} invariant are included in the gauge transformations \eqref{epsilonForm} that keep the background and the boundary conditions form-invariant. In other words, to identify the isometries, we demand now
\begin{align}
\label{eq:gaugetrafo0}
\delta_\epsilon A &= \extd\epsilon + \left[A,\epsilon\right]=0 & \delta_{\eb}\Ab &= \extd\eb + \left[\Ab,\eb\right]=0\ .
\end{align}
This means variations of the form \eqref{eq:angelinajolie} are not allowed anymore and we get restrictions for the functions $\epsilon_{L}$, $\epsilon_{W}$, $\eb_{L}$ and $\eb_{W}$ of the gauge generators \eqref{epsilonForm}. These are of the form
\begin{subequations}
  \label{eq:wedgerestriction}
  \begin{align}
0&=-\tfrac{k}{\pi}\eL^{(3)}(x)+\L'(x) \eL(x) +2 \L(x) \eL'(x)+2 \W'(x) \eW(x) +3 \W(x) \eW'(x)\\
0&= \tfrac{k}{\pi}\eW^{(5)}(x) + \tfrac{16 \pi}{k}\L(x) \L'(x) \eW(x)-2 \L^{(3)}(x) \eW(x)+\tfrac{16\pi}{k} \L(x)^2 \eW'(x) -9 \L''(x) \eW'(x) \nonumber\\
 &\quad -15 \L'(x) \eW''(x)-10 \L(x)\eW^{(3)}(x)+\frac{16}{3}\W'(x) \eL(x) +16 \W(x) \eL'(x)
  \end{align}
\end{subequations}
and are isomorphic for the barred sector after the replacement $\epsilon \to \eb$, $\mathcal{L} \to \Lb$ and $\mathcal{W} \to \Wb$. Since they are isomorphic, we shall only analyze the unbarred sector hereafter.

We restrict first to constant-valued $\L$ and set $\W=0$. Under this restriction, \eqref{eq:wedgerestriction} decouple $\eL$ and $\eW$, leading to
\begin{subequations}
  \label{eq:wedgeWzero}
  \begin{align}
    0&=\eL^{(3)}(x)-\tfrac{2 \pi}{k} \L \eL'(x)\\
    0&= \eW^{(5)}(x) -\tfrac{10 \pi}{k}\L\eW^{(3)}(x)+\tfrac{16\pi^{2}}{k^{2}} \L^2 \eW'(x)  .
  \end{align}
\end{subequations}
Depending on the sign of $\L$, the solution comes in three classes. For $\L>0$, the solution belongs to the hyperbolic class
\begin{subequations}
  \label{eq:modes}
  \begin{align}
    \eL(x)&= \sum_{n=-1}^{1} \eLn e^{n \omega x} & \eW(x)&= \sum_{n=-2}^{2} \eWn e^{n \omega x}.
  \end{align}
For $\L<0$, the solution belongs to the trigonometric class
  \begin{align}
    \eL(x)&= \sum_{n=-1}^{1} \eLn [ \cos (n \omega x + \varphi_L) ]
&\eW(x)&= \sum_{n=-2}^{2} \eWn [ \cos(n \omega x + \varphi_W) ]
  \end{align}
\end{subequations}
where $\omega=\sqrt{2 \pi \L / k}$.
For $\L=0=\W$, the solution belongs to the rational class
  \begin{align}
\label{eq:wedgelif}
    \eL(x)&= \sum_{n=-1}^{1} \eLn x^{n+1}  & \eW(x)&= \sum_{n=-2}^{2} \eWn x^{n+2} \ .
  \end{align}
Together with the barred sector, they form 16 linearly independent solutions to the isometry conditions \eqref{eq:gaugetrafo0}. Plugging these variations into the formulas for the charges, (\ref{charges}), we find that these isometries correspond to the modes $\L_{0,\pm 1}$, $\Lb_{0,\pm 1}$, $\W_{0,\pm 1,\pm 2}$ and $\Wb_{0,\pm 1,\pm 2}$\footnote{Fields with weight $h$ are expanded as $\phi(z)=\sum_{n \in \mathbb{Z}} \phi_{n} z^{-n-h}$. The quantity $\L(x)$  is  of weight $2$ and $\W(x)$ is of weight $3$.}, which generate exactly the wedge algebra $\mathfrak{sl}(3,\R)\oplus\mathfrak{sl}(3,\R)$ of our ASA $\W_{3} \oplus \W_{3}$. See again \eqref{eq:W3}.

Which class of $\L$ corresponds to the Lifshitz vacuum depends on the global structure of the $x$-space.
Hereafter, we shall assume that $x$ takes values in $\mathbb{R}$ with no periodicity built in, so that the dual CFT$_2$ is defined on a plane and not on a cylinder. In that case, the hyperbolic and trigonometric classes above that arise for $\L\neq 0$
have essential singularities at $|x|\to\infty$. Thus, the rational class \eqref{eq:wedgelif} is the only possible choice. In this class, the Lifshitz background \eqref{backgroundLineEl}, \eqref{backgroundSpin3} is the unique $\mathfrak{sl}(3,\R)\oplus\mathfrak{sl}(3,\R)$ invariant state in the theory on the plane and therefore corresponds to the ground-state on the CFT$_2$ side.

We recognize that the $z=2$ Lifshitz isometries \eqref{eq:iso} of our vacuum state are nothing but restriction to the 8 generators \eqref{eq:wedgelif},
\begin{subequations}
\label{eq:Lifgenerators}
\begin{align}
 \xi_\mathbb{H}:&&  \eWnn{-2} &= \frac49 & \ebWnn{-2} &= 1 & \textrm{remaining\;} \eLn, \eWn,\ebLn,\ebWn &= 0 \\
 \xi_\mathbb{P}:&&  \eLnn{-1} &= 1 & \ebLnn{-1} &= 1 &  \textrm{remaining\;} \eLn, \eWn,\ebLn,\ebWn &= 0 \\
 \xi_\mathbb{D}:&&  \eLnn{0} &=-1 & \ebLnn{0} &= -1 & \textrm{remaining\;} \eLn, \eWn,\ebLn,\ebWn &= 0,
\end{align}
\end{subequations}
by virtue of the relation between gauge symmetries and diffeomorphisms, $\epsilon = \xi^\mu A_\mu$, $\eb=\xi^\mu \Ab_\mu$ \cite{Witten:1988hc}. The gauge parameters $\eWnn{-2}$, $\eLnn{-1}$ and $\eLnn{0}$ generate the fluctuations $\W_{-2}$, $\L_{-1}$ and $\L_0$, respectively. With the identification $\W_{-2}\leftrightarrow \mathbb{H}$, $\L_{-1}\leftrightarrow \mathbb{P}$, $\L_0\leftrightarrow \mathbb{D}$ and the use of \eqref{eq:W3}, it becomes obvious that, with the barred sector included, we have the isometry subalgebra $\mathfrak{lif}(2,\R)$. 

So, the situation goes as follows. While one might naively expect due to diffeomorphism invariance that the Lifshitz spacetime should be invariant just under the Lifshitz isometry algebra $\mathfrak{lif}(2, \mathbb{R})$ \eqref{eq:Lalgebra}, we actually find an enhancement of it to the full wedge algebra $\mathfrak{sl}(3,\R)\oplus\mathfrak{sl}(3,\R)$. Crucially, the idea is that such enhancement is purely a phenomenon of higher-spin gauge symmetry: the higher spin gauge transformations mix the spin-3 background with the spin-2 metric background, and consequently the Lifshitz invariance gets enhanced to the full spin-3 higher-spin invariance.

It is worth noting that the isometry conditions \eqref{eq:wedgerestriction} admit no solutions that have only the Lifshitz isometries \eqref{eq:Lifgenerators}. 
Thus, any state in our theory must be either more symmetric (like our $z=2$ Lifshitz vacuum) or less symmetric. A simple class of examples for states with less isometries than the vacuum (but more than Lifshitz isometries) are configurations with fluctuations only in one sector, $\L=0=\W$  and $\ebLn=\ebWn=0$ which are invariant under a single $\mathfrak{sl}(3,\R)$ (and similarly for configurations with $\Lb=0=\Wb$ and $\eLn=\eWn=0$). A simple class of examples for states with less isometries than Lifshitz are solutions with constant $\L \neq 0$ and $\W \neq 0$. Restricting $\eL(x)$ and $\eW (x)$ to the modes \eqref{eq:wedgelif}, we only get isometries for nonzero $\eLnn{-1}$ and  nonzero $\eWnn{-2}$ whereas the other modes have to vanish. The situation is the same when either $\L$ or $\W$ vanishes (for $\W=0$ this can be seen from \eqref{eq:modes}, where only the constant modes are finite polynomials). The remaining two nonzero modes generate the commuting subalgebra $[\mathbb{H},\mathbb{P}]=0$ (with the identifications given above) of $\mathfrak{sl}(3,\R)$. We expect some of these states are black holes and conical surpluses, but we have not investigated them yet.

Although the isometry algebras are the same, the Lifshitz background and the AdS$_3$ background are different configurations, not related in any equivalent ways, such as regular gauge transformations. One way of showing this follows from the fact that the Lifshitz background is not invariant under the time-reversal ${\cal T}$. In terms of the connections, this is evident from (\ref{background}). In terms of higher spin fields, this is evident from (\ref{backgroundSpin3}). 
By contrast, the AdS$_3$ background is maximally symmetric and conserves ${\cal T}$. 
Furthermore, the theories themselves differ, as our Lifshitz boundary conditions are not ${\cal T}$-invariant, while the usual AdS  boundary conditions are.

Another way of showing that we are not describing AdS$_3$ holography in disguise is to try to explicitly construct a gauge transformation between the two theories. In other words, we want to find  $g$ and $\gb$ that fulfill
\begin{subequations}
\label{eq:6}
  \begin{align}
    A_{\rm{AdS}}&=g^{-1}\, A_{\rm{Lif}} \, g+ g^{-1}\extd g\\
    \Ab_{\rm{AdS}}&=\gb^{-1}\, \Ab_{\rm{Lif}}\, \gb+ \gb^{-1}\extd \gb,
\end{align}
\end{subequations}
where the AdS$_3$ is given by
\begin{subequations}
\label{eq:7}
  \begin{align}
    A_{\rm{AdS}}&= b^{-1} \left( \Lt_{+1} + \frac{2 \pi}{k} \L_{\mathrm{AdS}}(x^{+}) \Lt_{-1} -\frac{\pi}{2 k}  \, \W_{\mathrm{AdS}}(x^{+}) \Wt_{-2} \right)b \extd x^{+} +b^{-1}\extd b\\
    \Ab_{\rm{AdS}}&=-b \left(\Lt_{-1}+ \frac{2 \pi}{k} \, \Lb_{\mathrm{AdS}}(x^{-}) \Lt_{+1} -\frac{\pi}{2 k}  \, \Wb_{\mathrm{AdS}}(x^{-}) \Wt_{+2}\right)b^{-1} \extd x^{-}+b \extd b^{-1}.
  \end{align}
\end{subequations}
By $A_{\rm{Lif}}$ and $\Ab_{\rm{Lif}}$ we mean our usual connection given by \eqref{background} and \eqref{strictBC} (with the Lif subscript added for clarity).
To show equivalence we need to identify
\begin{subequations}
  \begin{align}
    \label{eq:3}
    \L_{\mathrm{Lif}}(x) &\Leftrightarrow \L_{\mathrm{AdS}}(x+t)&     \Lb_{\mathrm{Lif}}(x) &\Leftrightarrow \Lb_{\mathrm{AdS}}(x-t)\\
    \W_{\mathrm{Lif}}(x) &\Leftrightarrow \W_{\mathrm{AdS}}(x+t)&    \Wb_{\mathrm{Lif}}(x) &\Leftrightarrow \Wb_{\mathrm{AdS}}(x-t).
  \end{align}
\end{subequations}
A change of coordinates which transforms both $x^{+}=x+t$ and $x^{-}=x-t$ simultaneously to $x$ would require a singular diffeomorphism, and thus is not allowed. We therefore conclude that we are looking at a different theory than AdS$_3$. This also corroborates the fact that the Lifshitz critical point does not preserve the time-reversal ${\cal T}$. 

In conclusion, while the $z=2$ Lifshitz isometries \eqref{eq:iso} are part of the isometries \eqref{eq:wedgelif} of our Lifshitz vacuum, they get enhanced to the same set of isometries as in AdS$_3$, $\mathfrak{sl}(3,\R)\oplus\mathfrak{sl}(3,\R)$, due to the presence of a non-trivial spin-3 background field \eqref{backgroundSpin3}.

\section{Discussion}\label{concl}

In this paper, we established nontrivial results regarding a higher-spin realization of Lifshitz critical system in $(1+1)$ dimensions. We showed that, in higher-spin realizations, the anisotropic Lifshitz critical point actually exhibits isotropic and extended conformal invariance. Below, we highlight implications as well as several important issues for future studies. 

Firstly, it is interesting that higher-spin gravity offers a simple holographic realization of the Lifshitz spacetime. The higher-spin approach has advantages as compared to previously studied proposals. In particular, higher-spin theory, after Higgsing, is believed to be UV-completed to string theory (though precise details are not yet understood). Extending our construction to matter-coupled and supersymmetric higher-spin theories in $(2+1)$ dimensions \cite{Henneaux:2012ny} and to theories in higher dimensions is an interesting direction for future study. Explicit construction of dual Lifshitz field theory in $(1+1)$ dimensions and realization of the ${\cal W}_3 \oplus {\cal W}_3$ symmetry algebra is much sought for.

Secondly, it would be interesting to reassess various confusing results concerning the Lifshitz spacetime in Einstein gravity from the viewpoint of the higher-spin gauge symmetry. More specifically, Lifshitz holography seemed to lead to pathologies such as naked singularities and difficulties with embedding into string theory \cite{Copsey:2010ya,Horowitz:2011gh}. These issues may be absent in higher-spin realizations of Lifshitz spacetimes. Foremost, the metric is not higher-spin gauge invariant. For example, higher-spin gauge transformations can change the number of event horizons extracted from the metric \cite{Ammon:2011nk}. While higher-spin gravity in three dimensions is a topological theory, the topological invariants are not those of the manifold endowed with the metric induced by the spin-2 field. Rather, we should consider holonomies, which contain the gauge invariant information and can be used to determine global structures such as event horizons and singularities \cite{Gutperle:2011kf}. It is perhaps not surprising that the obvious isometries of the Lifshitz metric \eqref{Lifshitz3} can get (and indeed are) extended to the isometries of AdS$_3$ within spin-3 gravity.

Thirdly, we leave for future works the construction of the CFT$_2$ holographically dual to the Lifshitz background \eqref{background} with the boundary conditions \eqref{strictBC}. The asymptotic symmetry algebra $\W_3\oplus\W_3$ (see section \ref{se:4}) and its central charge turn out isomorphic to those associated with spin-3 gravity in asymptotically AdS$_3$ spacetime. We pointed out that the two systems are still distinguishable by the time-reversal ${\cal T}$. Related to this, we note that the CFT dual to the Lifshitz theory need not be modular invariant. As such, the presence of two left-moving sectors does not seem to pose a problem as it would for a theory defined on a torus. Moreover, the allowed classical saddle points could differ between the two theories. 

Fourthly, an exhaustive classification of admissible boundary conditions needs to be done. In addition to the boundary conditions discussed in section \ref{strict}, there is also a relaxed set of consistent boundary conditions for the same background (\ref{background}). These boundary conditions result in four towers of asymptotic charges in each of the unbarred and barred sectors and are explicitly shown in \cite{Prohazka:2013}. Unfortunately, the resulting asymptotic symmetry algebra is quite complicated. While the number of charges suggests that the algebra might be the Polyakov--Bershadsky algebra $\W_3^{(2)}\oplus\W_3^{(2)}$, verifying this conjecture would require a complicated non-linear field redefinition which we have not yet succeeded in finding. It is also possible to consider mixed boundary conditions, with the unbarred sector having the strict boundary conditions of section \ref{strict} and the barred sector having looser boundary conditions (or vice-versa), as both boundary conditions are based on the same background (\ref{background}).

We close with a discussion concerning the recent work of \cite{Gutperle:2013oxa}. The authors analyzed a set of higher spin boundary conditions leading to a Lifshitz sub-algebra within the asymptotic symmetry algebra, and presented its own set of puzzles. Their charges are finite but not conserved. Note that the non-conservation does not invalidate the symmetry algebra, since it can arise from the choice of a different charge basis, as explained in \cite{Bunster:2014mua}.
Interestingly, their boundary condition preserving gauge transformations, Eqs.~(3.31) and (3.32) in \cite{Gutperle:2013oxa}, also lead to a $\W_3$ algebra, as pointed out in \cite{Perez:2014pya}. 
In fact, their field configurations turn out to be a special case of a general class of solutions of spin-3 gravity in the presence of chemical potentials \cite{Compere:2013nba,Henneaux:2013dra}. In the conventions of \cite{Bunster:2014mua} [see their Eqs.~(3.7)-(3.11)] the relevant choice of chemical potentials is $\eta^\pm=1$ and $\xi^\pm=0$. 

Built upon their work and ours, we may put forth a conjecture for generic higher-spin Lifshitz holography that the asymptotic symmetry algebra gets ubiquitously enhanced from the infinite-dimensional Lifshitz algebra, for instance the one presented in appendix B of \cite{Compere:2009qm}, to a class of $\W$-algebras. Our present work can be viewed as a concrete realization of the conjecture for a specific case of spin-3 gravity.

\acknowledgments

We thank Hamid Afshar, Glenn Barnich, Andrea Campoleoni, St\'ephane Detournay, Michael Gutperle, Yu Nakayama, Yaron Oz, Eric Perlmutter,  Radoslav Rashkov, Max Riegler, Simon Ross, Jan Rosseel, Adam Schwimmer and Ricardo Troncoso for discussions. 

This work was supported by the START project Y~435-N16 of the Austrian Science Fund (FWF) and by the FWF projects I~952-N16 and I~1030-N27. DG acknowledges support by FAPESP and a BMWF--MINCyT bilateral cooperation grant, \"OAD project AR 09/2013. SJR acknowledges support by the National Research Foundation of Korea (NRF-MSIP) grants 2005-0093843, 2010-220-C00003 and 2012-K2A1A9055280. 
We thank the Galileo Galilei Institute for Theoretical Physics (MG, DG, SP), the Weizmann Institute (MG, SJR), the Asia Pacific Center for Theoretical Physics (DG, SJR), the Centro de Ciencias de Benasque (DG), ABC do Federal University (DG), ICTP SAIFR (DG, SJR), the Academy of Sciences of the Czech Republic (MG), the Universit\'{e} Libre de Bruxelles (MG, DG), the University of Buenos Aires (DG), and IAFE (DG) for their hospitality, and the INFN and the COST action MP1210 ``The String Theory Universe'' (MG) for partial support.

\appendix

\section{Spin-3 generators and their traces}\label{app:A}
The $\mathfrak{sl}(3,\R)$ algebra
\begin{subequations}
\label{eq:sl3}
  \begin{align}
    \left[\Lt_n,\,\Lt_m\right]&=(n-m)\,\Lt_{n+m} \\
    \left[\Lt_n,\,\Wt_m\right]&=(2n-m)\,\Wt_{n+m} \\
    \left[\Wt_n,\,\Wt_m\right]&=\sigma\,(n-m)(2 n^2 + 2 m^2 - n m -8)\, \Lt_{n+m}
  \end{align}
\end{subequations}
is generated by the following representation (with $\sigma=-\tfrac{3}{16}$)
\begin{subequations}
  \begin{align}
    \Lt_{+1} &= \left(\begin{array}{ccc}
        0&0&0\\
        -\sqrt{2}&0&0\\
        0&-\sqrt{2}&0
      \end{array}\right)
    & \Lt_0 &= \left(\begin{array}{ccc}
        1&0&0\\
        0&0&0\\
        0&0&-1
      \end{array}\right)
    & \Lt_{-1} &= \left(\begin{array}{ccc}
        0&\sqrt{2}&0\\
        0&0&\sqrt{2}\\
        0&0&0
      \end{array}\right)\\
    \Wt_{+2} &= \left(\begin{array}{ccc}
        0&0&0\\
        0&0&0\\
        3&0&0
      \end{array}\right)
    & \Wt_{+1} &= \left(\begin{array}{ccc}
        0&0&0\\
        -\frac{3}{2\sqrt{2}}&0&0\\
        0&\frac{3}{2\sqrt{2}}&0
      \end{array}\right)
    & \Wt_0 &= \left(\begin{array}{ccc}
        \frac{1}{2}&0&0\\
        0&-1&0\\
        0&0&\frac{1}{2}
      \end{array}\right)\\
    \Wt_{-1} &= \left(\begin{array}{ccc}
        0&\frac{3}{2\sqrt{2}}&0\\
        0&0&-\frac{3}{2\sqrt{2}}\\
        0&0&0
      \end{array}\right)
    & \Wt_{-2} &= \left(\begin{array}{ccc}
        0&0&-3\\
        0&0&0\\
        0&0&0
      \end{array}\right).
    &&
  \end{align}
\end{subequations}
The non-vanishing traces of bi-linears in generators are listed below.
\begin{subequations}
  \begin{align}
    \tr(\Lt_{+1}\Lt_{-1}) &= -4 & \tr(\Lt_0\Lt_0) &= 2 & &\\
    \tr(\Wt_{+2}\Wt_{-2}) &= 9 & \tr(\Wt_{+1}\Wt_{-1}) &= -\frac{9}{4} & \tr(\Wt_0\Wt_0)
    &= \frac{3}{2}
  \end{align}
\end{subequations}

\section{$\boldsymbol{\W_3}$ algebra at finite central charge}\label{app:B}

The variation \eqref{eq:angelinajolie} derived in the main text translates into corresponding Poisson brackets \eqref{shortcut} that generate the ASA. Replacing Poisson brackets by commutators and introducing modes for the generators then leads to the ASA in the classical (large central charge) limit. For finite values of the central charge, normal ordering effects shift some of the structure functions in the algebra. The final result of this analysis is the $\W_3$ algebra at finite central charge (first introduced in \cite{Zamolodchikov:1985wn} and reviewed in \cite{Bouwknegt:1992wg})
\begin{subequations}
 \label{eq:W3}
\begin{align}
 [\L_n,\,\L_m] &= (n-m)\,\L_{n+m} + \frac{c}{12}\,(n^3-n)\,\delta_{n+m,\,0} \\
 [\L_n,\,\W_m] &= (2n-m)\,\W_{n+m} \\
 [\W_n,\,\W_m] &= (n-m)(2n^2+2m^2-nm-8)\,\L_{n+m} +\frac{c}{12}\,(n^2-4)(n^3-n)\,\delta_{n+m,\,0}  \nonumber \\
&\quad + \frac{96}{c+\tfrac{22}{5}}\,(n-m) \, \Lambda_{n+m}
\label{eq:quadratic}
\end{align}
\end{subequations}
where
\begin{equation}
  \label{eq:lambda}
\Lambda_{n}=  \sum_{p\in\mathbb{Z}} :(\L_{n-p}\L_p): -\frac{3}{10} (n+3) (n+2) \L_{n} \ .
\end{equation}
For the corresponding analysis in the AdS case, see \cite{Henneaux:2010xg,Campoleoni:2010zq}.
The generators split into the towers of Virasoro generators $\L_n$ and of spin-3 generators $\W_n$ with integer $n$. For large values of the central charge $c$, the quantum shift of $\tfrac{22}{5}$ in the denominator in the last line of \eqref{eq:W3} is negligible. The wedge algebra $\mathfrak{sl}(3,\R)\oplus\mathfrak{sl}(3,\R)$ [see \eqref{eq:sl3}] is recovered by first restricting to the wedge  modes $\L_{0,\pm 1}$ and  $\W_{0,\pm 1,\pm 2}$  and then taking the $c \rightarrow \infty $ limit.

\end{document}